# Reduced Graphene Oxide and Its Natural Counterpart Shungite Carbon

Elena F. Sheka

*Russian Peoples' Friendship University of Russia*
*Moscow, 117198 Russia*

sheka@icp.ac.ru

**Abstract.** Large variety of structure and chemical-composition of reduced graphene oxide (RGO) is explained from a quantum-chemical standpoint. The related molecular theory of graphene oxide, supported by large experience gained by the modern graphene science, has led the foundation of the concept of a multi-stage graphene oxide reduction. This microscopic approach has found a definite confirmation when analyzing the available empirical data concerning both synthetic and natural RGO products, the latter in view of shungite carbon, suggesting the atomic-microscopic model for its structure.

**Key words:** reduced graphene oxide, technical graphene, shungite carbon, graphene oxide reduction, quantum-chemical approach, molecular theory

## 1. Introduction

According to the judgment of competent experts [1], the modern graphene technology can be divided into two independent domains, namely, low-performance (LP) and high-performance (HP) ones. The first includes a wide spectrum of practical applications based on graphene nanomaterials. The characteristic products of this domain are modified polymer and other composites, sensors and sensor screens, roll-up electron paper, organic light-emitting diodes, and so forth. The products of the second domain are based on micro- or larger sized one- or multilayer graphene sheets and represent electron devices, such as high-frequency, logic, and thin film field-effect transistors. This *de facto* division of the graphene technology into two types results from the molecular–crystalline dualism of the graphene nature and the technical implementation of its unique chemical and physical properties rather than from the simplification of operation with complex technologies [2].

  The objective reasons of postponing the graphene HP technology [1, 3] up to 2030 are the serious problems related to the development of technologies intended for mass production of micro and macrosized crystalline graphene sheets, which is complicated by the high cost of this material [4].The implementation of the LP technology is more successful. The active efforts of numerous chemist teams solved the problem of mass production of the required technological material, namely, technical graphene. This material is the end product of a complex redox technological cycle, involving fragmentation of graphite to nanoparticles followed with the particle oxidation and formation of graphene oxide (GO) and completed with the GO reduction. In all the cases, structural analysis demonstrates well-pronounced non-flatness of GO molecules and almost entire restoration of the flatness of the basal plane of reduced graphene oxides (RGO). Therefore, RGO is mentioned as graphene in many works. However, in contrast to the technological materials used to date (which are usually rigorously standardized in chemical composition and structure), the standardization of technological graphene seems to be impossible, since this term covers a very wide set of substances, which represents various oxyhydride polyderivatives of graphene nano- and micro-molecular sheets and-or-molecules. All the substances of this class are characterized by the flatness of their carbon skeletons but differ by the chemical groups that terminate dangling bonds along their perimeter [2, 5]. Evidently, the structure and chemical composition of RGO can change at each of the three stages of chemical synthesis mentioned above. The latter results in many versions of the chemical composition as well as shape and structure of synthesized RGOs, which is being actively discussed [6]. For example, the residual oxygen concentration, which is a very important parameter of the material, can differ 20 times in different productions.

## 2. Reasons for an intrinsic variability of RGO

Graphene chemistry drastically differs from the conventional molecular one and presents a very large and complicated domain related to substances with spatially distributed targets (see review [7] and references therein). Nevertheless, despite a great variety, morphologically, graphene-based (derivative) molecules can be divided into three groups: (i) verily graphene molecules (VGMs) presenting pieces of flat honeycomb sheets with non saturated dangling bonds of edge atoms; (ii) framed graphene molecules that are the above VGMs with saturated dangling bonds in the circumference area (FGMs or CFGMs); and bulk graphene molecules (BGMs) with chemical addends enveloping the whole body of the carbon skeleton. Particular examples of these three groups can be easily found among the extended collection of graphene chemicals [6, 8-10]. The first group should be attributed to the pristine molecules while two other are related to the VGMs polyderivatives. Evidently, the division is quite expected and just reveals the unique two-zone feature of the chemical activity of pristine VGMs that governs the formation of any of their derivatives [2, 5,7].

While VGMs and BGMs are completely different, including both chemical compositions and the carbon skeleton structure, FGMs show not only the difference with the two groups, but a commonality as well. With respect to VGMs, the FGMs conserve chemically untouched basal plane that even though disturbed maintains some flatness thus keeping graphene-like style. On the other hand, a polyderivative origin joins FGMs and BGMs, which makes them both to be different from VGMs.

Accordingly to this three-group division, the available graphene materials are evidently of three categories. Materials of the first category present graphene crystals in the form of macro- micrometer-size perfect one-atom thick sheets and can be attributed to the VGM group. Actually, real graphene sheets are FGMs however their large size and relatively small number of edge carbon atoms allow for neglecting the frame influence since the main actions concerning the material occur far from the sheet edges. Materials of the second category are related to 'technical graphene'. Nowadays they, originally produced from GOs, are presented by highly variable RGOs and should be attributed to micrometer- and/or nanometer-size

FGMs. Since FGM formation is always connected with edge carbon atoms, each of which are characterized by high chemical reactivity of more than 1 e per atom [2, 5, 7], the material stabilization under ambient conditions imperatively requires the activity inhibition due to which framing of edge atoms by chemical addends must be completed, not leaving even a single edge atom not attached. There are some nuances related to if one- or two-valence addend is attached to the relevant carbon atoms. The issue was considered in [5, 11] on the examples of graphene oxidation and hydrogenation thus explaining a large variety of produced RGO products with respect to the chemical composition of their framing. Micrometer- and/or nanometer-size sheets of graphene oxides, hydrides, fluorides, and so forth present materials of the third category and can be marked as 'modified graphene'. Since they represent BGMs, the extent of the pristine VGMs modification in this case is high. It must exceed the saturation level characteristic to FGMs but should not be imperatively completed, once possessing a large variety.

Besides the names, the three graphene materials drastically differ by their behavior because of deeply rooted discrepancy between them. Thus, the transformation of perfect graphene into materials of categories 2 and 3 is well mastered and developed using a large spectrum of different chemical technologies but absolutely irreversible. This is mainly due to the loss of the integrity of pristine graphene sheets caused by their cracking under chemical treatment. This is one of more other reasons why any device made of perfect graphene must be reliably protected from the environmental chemical attacks. As for technical and modified graphenes, they are much more resistant and even a reversible transformation between them is partially possible. However, each cycle of such transformation is followed by reducing size of the relevant sheets. Energy-lite deposition on and/or removing of chemicals from the basal atoms of the carbon skeleton greatly favors the action to be performed.

Since the reduction of massively produced GOs is the main way of producing technical graphene (RGO) now, we shall begin the RGO consideration by looking the inherent connection between GO and RGO from the points of the molecular theory of graphene. Starting the heralded consideration, let us try to answer the following questions:

1. Is it possible to liberate any GO from oxygen containing groups (OCGs) completely and if not, to what extent?
2. Is it possible to return a planar honeycomb structure to drastically deformed and curved carbon skeleton structure of GOs?

Obviously, RGO chemical composition and morphology are tightly connected with those of pristine GO ones thus relating to the latter when removing all OCGs from the sheet basal plane. On this background, let us address the (5, 5) GO molecule (GO X) described in details in [5] and reproduced in Fig. 1a. It was computationally synthesized in the course of the stepwise oxidation of the pristine (5, 5) nanographene (NGr) molecule in the presence of three oxidants, such as O, OH, and COOH. Since the removing of each of these oxidants as well as their attachment are characterized by the same coupling energy, it is evident that the per-step coupling energy (PCE) distribution presented in Fig. 1b describes not only the attachment of the groups to either basal plane atoms (curves 1 and 2 in Fig. 1b) or edge atoms in the circumference area (curve 3) on the way from the pristine (5, 5) NGr molecule to GO X, but removing the oxidants from the latter towards (5, 5) RGO. As evidences from the figure, since the PCE values are very different, the reduction should be obviously expected as multistage or multimode one. Actually, the oxygen atoms located in the basal plane of GO X molecule and forming mainly epoxy groups with carbon atoms (within the rose shading) should be removed first. The corresponding (5, 5) RGO molecule is shown in Fig. 1c. This apparently happens at the first stage of the real reduction and may present the final state of the reduction procedure when the latter is either short-time or not very efficient, once to be attributed to a mild one. The corresponding mass content of the obtained (5,5) RGO molecule is given in Table 1.

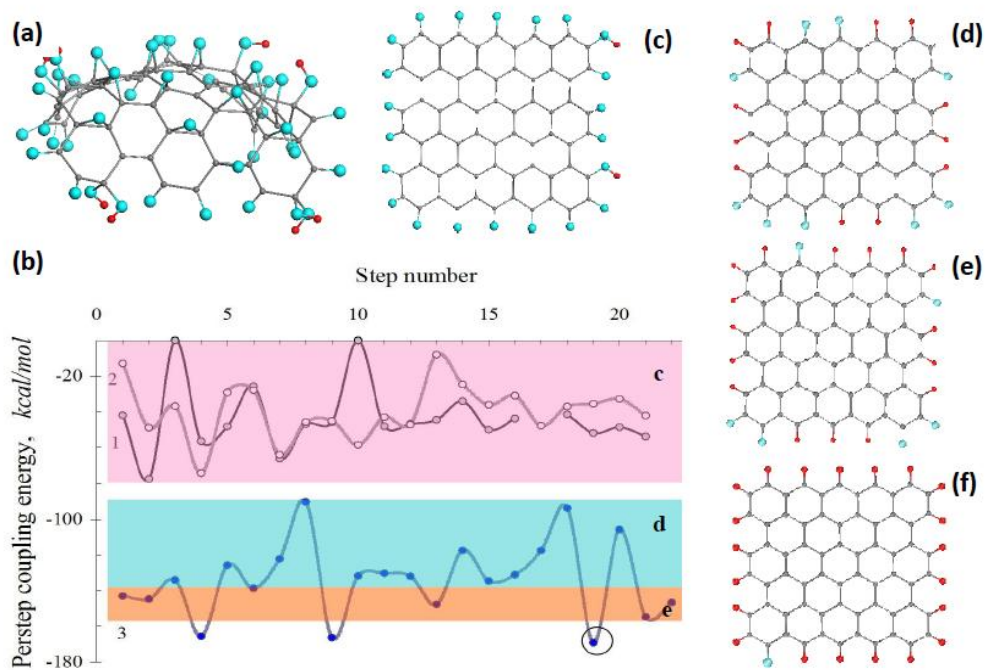

**Figure 1**. (a) Equilibrium structure of ~1 nm (5, 5) GO sheet (GO X according to [5]) corresponding to one-side oxidation of the pristine (5, 5) NGr molecule. (b) Per-step coupling energies related to the one-side oxygenation of the (5, 5) NGr molecule: O- and OH-attachments to the basal plane (curves 1 and 2, respectively) and the combination of O and OH attachments in the circumference (curve 3) [5]; the circled point corresponds to the formation of carbonyl unit on the RGO circumference with the largest coupling energy. (c) Model (5, 5) RGO sheet corresponding to the first stage of GO X deoxygenating that affects the atoms in the basal plane of the GO X sheet only (*c* area in (b)) (mild reduction). (d) and (e) Model (5, 5) RGO sheets corresponding to a medium and hard reduction of GO X in the framework of the blue d and cream e shaded zones in (b), respectively. (f) Model (5, 5) RGO sheet corresponding to circled point in (b). Color code: gray, blue and red balls present carbon, oxygen, and hydrogen atoms, respectively. AM1 UHF calculations.

When the reduction occurs during long time or under action of strong reductants, it may concern OCGs located at the RGO circumference. Such two-step reduction of a pristine GO has been convincingly fixed in practice [12]. The second-step reduction faces the following peculiarities. First, due to a waving character of the PCE

Table 1. Chemical composition and mass content of differently reduced (5, 5) RGOs

| Atomic composition | Mass content, wt % | | | Remarks |
|---|---|---|---|---|
| | C | O | H | |
| $C_{66}O_{22}H_2$ $(C_6O_{0.54}H_{1.45})^*$ | 69,13 | 30,70 | 0,17 | Fig. 1c |
| $C_{66}O_9H_{13}$ $(C_6O_{0.54}H_{1.45})^*$ | 83,46 | 15,17 | 1,37 | Fig. 1d |
| $C_{66}O_6H_{16}$ $(C_6O_{0.54}H_{1.45})^*$ | 87.6 | 10.6 | 1.8 | Fig. 1e |
| $C_{66}O_1H_{21}$ $(C_6O_{0.09}H_{1.91})^*$ | 95.5 | 1.9 | 2.5 | Fig. 1f |
| Sh-rGO | 95.3-92.4** | 3.3-2.5** | 2.0-0.7*** | Exp. [13] |

*Averaged mass content per one benzenoid unit.

**The scatter in the data obtained from measurements in four points of one sample [13].

***The hydrogen content is difficult to determine due to which only the range is definitely fixed [13].

dependence with large amplitude from -90 kcal/mole to -170 kcal/mole, the second step reduction could be highly variable due to the fact that in practice the applied reduction protocol usually concerns the liberation of atoms with coupling energy that is restricted by the protocol conditions. Thus, restricting the energy interval to 30 kcal/mol (removing oxidants covered by blue shading in Fig.1b) results in remaining only 9 oxygen atoms instead of 22 in the first (5, 5) RGO sheet shown in Fig.1c. Second, since the liberation concerns edge carbon atoms, a release of one of them from oxygen or hydroxyl makes the atom highly chemically active, which imperatively requires the inhibition of its chemical activity. This work can be done by hydrogen atoms thus transforming the reduction into deoxygenation/hydrogenation procedure. The suggestion is well consistent with, first, a scrupulous analysis of C:O content of differently produced RGOs [6, 9] which convincingly evidences a strong dependence of the ratio from the reduction protocol in use, and, second, a pronounced hydrogen content detected in real RGO samples [13].

Coming back to Fig. 1b, we see that the reduction within blue zone halves the oxygen content and causes the emergence of a remarkable quantity of hydrogen. The mass content of the corresponding RGO molecule shown in Fig. 1d is given in Table 1. Further strengthening of the reduction, counted by lowering the number of remaining oxygen atoms, gradually decreases the oxygen content while increasing the hydrogen contribution. Consequently, the structures shown in Figs. 1c, e, and f might be attributed to RGOs obtained in the course of mild, medium, and hard reduction, respectively, thus presenting RGOs as a multi-stage series of different FGOHs.

Completing the answer to the first question posed above we can formulate it in the following form. The liberation of GOs basal plane from OCGs is always completed and results in the formation of FOHGs. The circumference area of the latter involves mainly oxygen and hydrogen atoms, individually attached to edge atoms of the carbon skeleton, since otherwise it would be impossible to keep a standard RGO interlayer distance of ~0.35±0.015 nm, instead of ≥ 0.7 nm accustomed for GO, that is a standard characteristic of the species [14]. Both the oxygen and hydrogen contents of RGO depend on the reduction protocol in use.

Addressing the second question, we can again refer to empirical evidence on keeping the interlayer distance in RGO at the level of ~0.35±0.015 nm, which is possible if only carbon skeletons of RGOs are considerably flattened. However, the $sp^2 \rightarrow sp^3$ reconfiguration of the FGM carbon skeleton is accompanied with huge deformation energy, compensation of which is needed when flattening the skeleton. The computational experiment can show if such a transformation is possible spontaneously without additional requirements. Thus, let us release GO X molecule from all the oxidant atoms just removing them from the output files and conserving the skeleton (core) structures therewith. The core structure of the molecule, shown on the top of Fig. 2a, was subjected to optimization. As seen in Fig. 2b, the core optimization fully restores the planar structure of the (5, 5) NGr molecule. The restoration is maintained if OCGs are removed from the basal plane only as seen on the bottom of Fig. 2b. The difference in the total energy of thus restored structures constitutes not more than a few tenths of a percent of the energy of both (5, 5) NGr and (5, 5) RGO molecules, the latter shown in Fig. 1c. Therefore, a drastic deformation of the GOs carbon skeletons is no obstacle for the restoration of the initial planar graphene pattern in spite of the large deformational energy of 874 kcal/mol incorporated in the deformed core. This is an exhibition of the extreme flexibility of the graphene structure that results in a sharp response by structure deformation on any external action, on the one hand, and provides a complete restoration of the initial structure, on the other.

**3. Rediscovery of shungite carbon in light of reduced graphene oxide**

Although many ways of generating single atomic layer carbon sheets have been developed, chemical exfoliation of graphite powders to GO sheets followed by deoxygenation/reduction to form chemically modified RGO has been so far the only promising route for bulk scale production [15, 16]. However, available technologies face a lot of problems among which there are low yield, the potential fire risk of GO and RGO when alkaline salt byproducts are not completely removed, a great tendency to aggregation, a large variety of chemical composition, and so forth. Additionally, chemical reduction usually requires toxic chemicals, several tedious batch steps, high temperatures and energies, and special instruments and controls for the preparation of RGO. Thus, an environmentally friendly preparation strategy is highly desired. In light of this, the existence of natural RGO is of the utmost importance. As if anticipating the future need for the substance, the Nature has taken care of a particular carbon allotrope in the form of well-known shungite carbon from deposits of carbon-rich rocks of Karelia (Russia) that strongly has been keeping secrets of its origin and the RGO-based structure. Just only recently has become clear that shungite carbon has a multilevel fractal structure based on nanoscale RGO sheets [17] that are easily dispersible in water and other polar solvents and presents a natural pantry of highly important raw material for the modern graphene technologies. This suggestion has been the result of a careful analysis of physico-chemical properties of shungite widely studied to date as well as was initiated by the knowledge, accumulated by both empirical and computational graphene chemistry, particularly related to graphene molecules hydrogenation, oxidation and/or reduction and its vision in the framework of the spin chemistry of graphene [7].

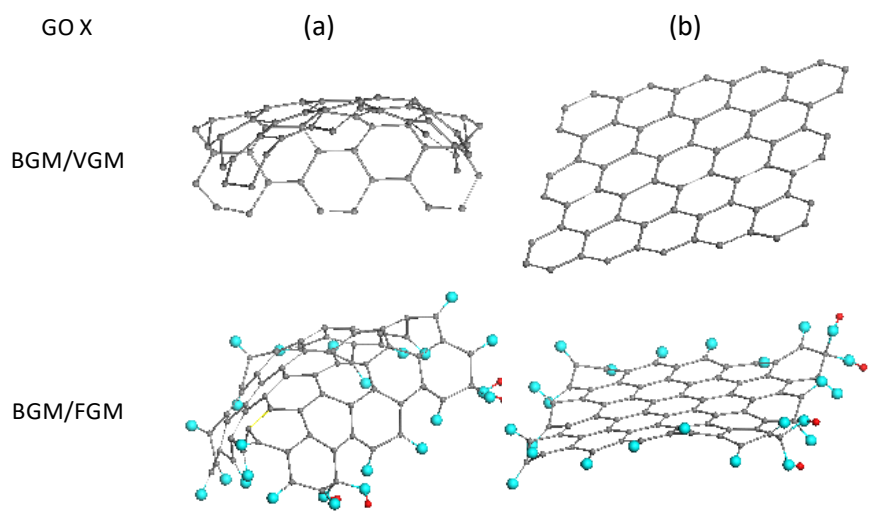

**Figure 2**. The core structures of graphene oxide GO X before (a) and after (b) BGM/VGM (top) and VGM/FGM (bottom) transformation. GO X molecule was obtained [5] for the pristine molecule basal plane accessible for the oxidants from the top only.

Shungite rocks, firstly found in the vicinity of town of Shun'ga in Karelia (Russia), are widely known and are in a large consumer's demand due to its unique physico-chemical and biomedical properties [18-20]. A lot of efforts have been undertaken to exhibit that the material, by ~98% pure carbon, presents a fractal structure of agglomerates consisting of nanosize globules [18], each of which presents a cluster of ~ 1 nm graphene-like sheets [19]. Figure 3 presents the current view on the microscopic structure of shungite carbon which, on the one hand, accumulates the long-term structural studies of shungite and, on the other, is based on the recent HRTEM studies [13] allowing to reveal the most important feature of the shungite structure at microscopic level. As seen in Figs. 3 (a-c), strips of bright spots of length from a fraction to a few nanometers form the ground of HRTEM images. The strips are the projection of planes composed of carbon atoms and oriented nearly parallel to electron beam. Fourier transformation applied to a selected area of the HRTEM image in Fig. 3 d produced Fourier diffractogram shown to the right of the area that is characteristic for a disordered graphite material with interlayer distance $d_{002}$=0.34 nm. At the same time, as seen in Figs. 3 (a-c) the graphite (graphene) elements form 4-7 layer stacks lateral dimension of which constitute 1.5-2 nm. Therefore, the data clearly show two first levels of the shungite carbon structure presented by individual graphene sheets of 1.5-2 nm in lateral dimension, which are the basic elements of the shungite carbon structure, and 4-7 layer stacks of the sheets. The formation of globes and their aggregates as the third and the fourth levels of the shungite structure, respectively, is convincingly confirmed by the fixation of two types of pores in the dense material, namely, small pores with linear dimensions of 2-10 nm and large pores of more than 100 nm in size [21]. The findings allowed suggesting that the shungite fractal structure is provided with aggregates of globular particles of ~6 nm in average size which are formed by the 4-7 layer stacks of the basic graphene fragments.

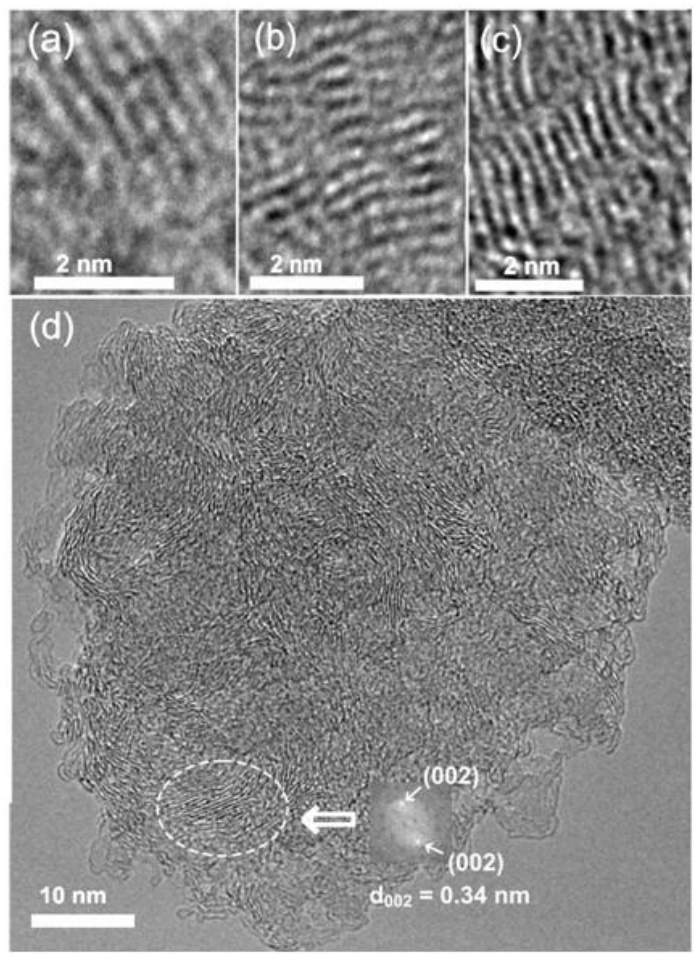

**Figure 3.** Dark field HRTEM images of fragments of atomic structure of shungite carbon: stacks of flat (a, b) and curved (c) graphene layers; (d) general view of shungite particles. Oval marks the area subjected to diffraction Fourier analysis results of which are given to the right. Adapted from Ref. 13.

The suggested structural model is based on flat basic fragments. Actually, such elements are characteristic for the real shungite structure as seen in Figs. 3 (a-c), where they dominate. However, in HRTEM images one can see curved graphene elements. In due time, the observation of these curved fragments laid the foundation of a fullerenic concept of shungite [22] which was not proved later on. The next suggestion concerning the curvature of basic elements connected the latter with the $sp^2s \rightarrow sp^3$ transformation of basal carbon atoms caused by the chemical modification within basal plane. It was rejected as well due to interlayer distance of 0.34 nm-0.38 nm maintained for stacks of curved fragments as well, which excluded the location of any chemical atom between the adjacent layers.

Remains to assume that the curvature of basic elements might be caused by the presence of other mineral inclusions, such as micro-nano-size particles of silica or different metals that always accompany shungite rocks. A large spectrum of traces of different metals, including Ni, Cu, Ag, Au, and some others, is known for shungite rocks [22] due to which the presence of the relevant nanoparticles is highly probable. The basic graphene fragments, enveloping metal nanoparticles, curve similarly to what was shown experimentally for a graphene sheet superimposed on a place of gold nanoparticles [23].

Modern experimental facilities allow for integrating high-resolution scanning transmission electron microscopy and point energy-dispersive X-Ray spectroscopy (EDS) thus providing a possibility to fix structural and chemical composition of the studied sample simultaneously. The averaged data related to the sample presented in Fig. 3d are listed in Table 1. As seen in the table, carbon and oxygen are the main element of the studied shungite carbon and constitute in average 95.5 ±0.6 *wt*% and 3.27±0.4 wt%. Hydrogen presence is in amount of 0.7 ± 0.15 wt%, which is well consistent with the relevant estimations of 0.5-1.5 wt% from the literature [24].

Presented data of the EDS analysis alongside with structural data given in Fig. 3 allow to suggest that the basic graphene element of shungite carbon represents a nanosize FGM molecule of the C:O:H=(95.5 ±0.6):( 3.27±0.4):( 0.5-1.5) composition. Since the lateral dimensions of FGM molecules shown in Fig. 1 is 1.3-1.4 nm, accordingly to data listed in Table 1 molecule (f) in the figure may be one of the most appropriate models of the shungite carbon basic element.

## 4. Shungite in view of graphene molecular chemistry of graphene

Since the Karelian shungite deposits of tens million of tons are unique and are not repeated at any other place of the Earth, it is highly challenging to trace how they were formed and why they present the deposits of natural RGO. The graphene-based basic structure provides a good reason to consider shungite carbon at the microscopic level by using high power of the modern empirical and theoretical molecular science of graphene. This approach allows not only to explain all the peculiarities of the shungite behavior, but also to lift the veil on the mystery of its origin.

Graphene-like structure of its basic element allows starting the story from that one of graphite. According to the privileged modern concept [25], graphite originates from precipitation of solid carbon from carbon-saturated C–O–H fluids. The transformation of carbonaceous matter involves structural and compositional changes of basic structural units in the form of aromatic lamellae (graphene sheets) and occurs in the nature in the framework of thermal or regional metamorphism that apart from temperature, involves shear strain and strain energy.

The graphitization is a longtime complicated process that, occurring during a geological scale of time, can be subjected to various chemical reactions. Tempo and character of the reactions are obviously dictated by the environment. It is quite reasonable to suggest that the aqueous environment at 300-400$^0$C, typical for Karelia area, under which the metamorphic evolution of carbonaceous matter occurs, is a dynamically changeable mixture of water molecules, hydrogen and oxygen atoms, hydroxyl and carboxyl radicals as well. The interaction of the carbonaceous matter, subjected to structural and compositional changes in the course of alignment of graphene lamellae and pore coalescence with this mixture accompanies the process. The most expected reactions concern hydration, hydrogenation, oxidation, hydroxylation, and carboxylation of the formed lamellae. At this point, it is important to note that any reaction of this kind primarily involves edge carbon atoms of the sheets. The addition, obviously, terminates the lamellae growth thus limiting the lateral dimensions of the formed graphene layers. Empirically, it has been repeatedly observed in the case of graphene oxide [8]. Therefore, since the above mentioned reactions start simultaneously with the deposit formation, their efficiency determines if either the formed graphene lamellae will increase in size (low efficient reactions) or the lamellae size will be terminated (reactions of high efficiency), which well correlates with a general conclusion made by Hoffmann when considering the chemistry role for nanoscience [26]: "too great stabilization will inhibit growth; too little stabilization will not prevent from collapse to the solid". Since large graphite deposits are widely distributed over the Earth, it might be accepted that aqueous environment of the organic-carbon-rich sediments generally does not provide both thermodynamic and kinetic conditions suitable for the effective termination of graphene lamellae in the course of the deposit graphitization. Obviously, particular reasons may change the situation that can be achieved in some places of the Earth. Apparently, this occurred in the Onega Lake basin, which caused the replacement of the graphite derivation by the shungite carbon formation. Some geologists reported on the correlation of the shungite formation with the increase of oxygen concentration in atmosphere that took place in 1.9- 2.1 Ga [27].

If chemical modification of graphene lamellae is responsible for limiting their size, the size limitation to ~1 nm should be sought in the relevant reaction peculiarities. First of all, one must choose the reaction that is preferable under the pristine graphenization conditions. Including hydroxylation and carboxylation into oxidation reaction, we must make choice among three of them, namely, hydration, hydrogenation, and oxidation. All the three reactions are well studied for graphene at molecular level both empirically and theoretically. The pristine graphene lamella is hydrophobic so that its interaction with water molecules is weak. Chemical coupling of a water molecule with graphene sheets can rarely occurs at the zigzag edge and is characterized by small coupling energy. Accordingly, water cannot be considered as a serious chemical reactant responsible for the chemical modification of the pristine graphene lamellae. Nevertheless, water plays extremely important role in the shungite fortune that will be shown later. As for graphene hydrogenation, it is a difficult task: the process usually involves such severe conditions as high temperatures and high pressure or employs special devices, plasma ignition, electron irradiation, and so forth that constitutes ~13 kcal/mol [17]. In contrast, graphene oxidation can apply for the role. The reaction has been thoroughly studied at different conditions (see reviews [15, 28-30] and references therein) and the achieved level of its understanding is high. The oxidation may occur under conditions that provide the shungite carbon derivation in spite of low acidity of the aqueous surrounding but due to long geological time and practically barrierless character of the reaction concerning additions of either oxygen atoms or hydroxyls to the graphene body [17]. As shown, oxidation causes a destruction of both pristine graphite and graphene sheets just cutting them into small pieces [29, 30]. This finding allows suggesting that shungite carbon sheets of ~1 nm in size have been formed in the course of geologically prolong oxidation of graphene lamellae derived from the graphenization of carbon sediments.

However, the oxide nature of shungite carbon strongly contradicts to empirical data related to fixed interlayer distance at the level of ~0.35 nm and low atomic percentage of oxygen in the carbon-richest shungite rocks. This contradiction forces to think about full or partial reduction of the preliminary formed GO occurred during geological process. The reduction procedure described in Section 2 shows that the natural environmental conditions are quite suitable for the GO reduction to occur. However, usually the synthetic GO reduction takes place in the presence of strong reducing agents that are not available in the natural environment of shungite. However, as was convincingly shown, the GO reduction can take place just in water, which only requires a much longer time for the process completion [31]. Evidently, the geological time of the shungite carbon derivation is quite enough for the reduction of pristine GOs in water to take place.

Basing on microscopical presentation of multistage GO reduction suggested above it is reasonable to suggest that each of the RGO molecules presented in Figs. 1(c-e) of 1.3x1.4 nm$^2$ could be a proper model of ~1 nm basic unit of shungite carbon and only O:H content may be a decisive factor in favor of the most appropriate one. According to empirically determined chemical composition of shungite carbon [13] and calculated chemical content listed in Table 1, the choice of model should be done in favor of the molecule shown in Fig. 1e. As other (5, 5) RGOs, it presents a FGM with a dominant contribution of framing hydrogen atoms. The latter finding raises a question of the hydrogen atoms origin in the shungite carbon environment. Actually, they partially may be due to spontaneous dissociation of water molecules at the surrounding temperature. However, one of the recent publications [32] concerning metal-induced GO reduction and hydrogenation, has forced to suggest the nascent hydrogen atoms play the main role in the hydrogen enrichment of RGO basic elements of shungite carbon. Actually, as shown, the nascent hydrogens willingly originate from the dissolving of metals in acidic media. However, a great deal of micro-nano metal inclusions accompanies shungite carbon while the aquatic surrounding is of weak acidity [13]. Evidently these two factors are highly favorable for reduction and/or hydrogenation of previously formed GO sheets of shungite carbon. Moreover, the nascent-hydrogen reaction causes an additional decrease of lateral dimensions of the final product thus contributing to stabilization of the basic RGO element size at the level of ~1 nm.

Assuming that RGO molecules of $C_{66}O_1H_{21}$ atomic composition in Fig. 1e present the first stage of the shungite carbon derivation, let us trace their path from individual molecules to densely packed shungite carbon shown in Fig. 3. Obviously, the path proceeds through successive stages of the sheets aggregation. Empirically was proven that aggregation is characteristic for synthetic GO and RGO sheets as well. Thus, infrared absorption [33] and neutron scattering [14, 34] showed that synthetic GO forms stacked turbostratic structures that confine water. Just recently, a similar picture was obtained from the studies of neutron scattering from two synthetic RGOs [34, 35] and shungite carbon [36, 37]. Neutron diffraction showed that the characteristic graphite interfacial distance $d_{002}$ constitutes, in average, ~0.69 nm in the case of GO and reduces to ~0.35 nm for RGO of both synthetic and natural origin, evidently indicating the recovery of the GO carbon skeleton planarity. The characteristic diffraction peaks of shungite are considerably broadened in comparison with those of graphite, which allows for estimating approximate thickness of the stacks of ~1.5 nm [37] that corresponds to five-six-layered RGO sheets. Figure 4a shows stacks of RGO nanosheets of $C_{66}O_1H_{21}$ atomic composition that form the secondary structure of shungite. Shungite confines ~4wt% of water but none of the water molecules can be retained near the RGO basal plane once kept in the vicinity of oxygen framing atoms (see Fig. 4b). The finding is well correlated with the RGO short-packed stacked structure leaving the place for water molecules confinement in pores formed by the stacks. The neutron scattering study has proven the pore location of retained water in shungite [37]. The water can be removed by mild heating thus dividing shungite carbon into 'wet' and 'dried'.

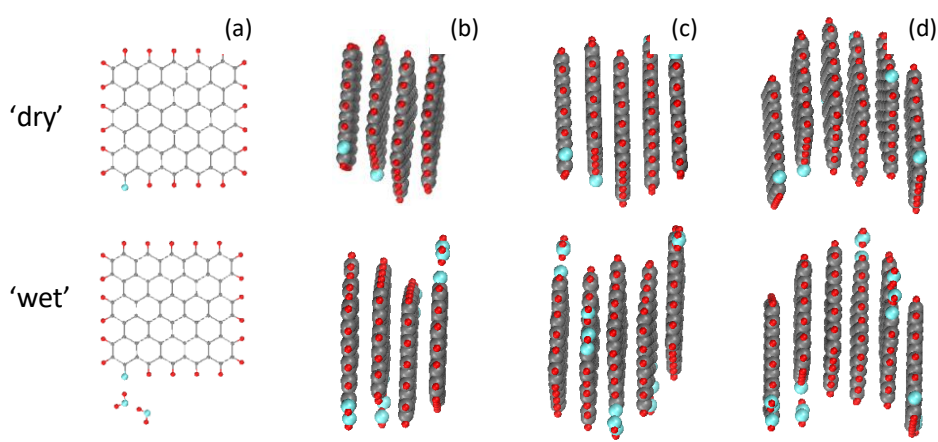

**Figure 4**. Components of 'dry' and 'wet' shungite carbon. (a) Equilibrium structure of ~1 nm (5, 5) RGO sheets of the $C_{66}O_1H_{21}$ chemical composition, 'dry' (top) and 'wet' (bottom), respectively. (b) - (d) Arbitrary models of four-, five-, and six-layer stacks of the relevant RGO sheets with 0.35 nm interlayer distance. See atom color code in the caption to Fig.1.

Further aggregation combines the stacks in globules, a planar view of one of possible models related to 'dry' shungite is presented in Fig. 5a. The globules of a few nm in size present the third stage of the shungite carbon structure [18]. Accordingly, the globe model of 'wet' shungite may look as presented in Fig. 5b. Irregular distribution of stacks in space causes the formation of interglobular pores, linear dimensions of which are comparable with those of stacks. Actually, one of the linear sizes of the pores formed by RGO nanosheets is determined by linear dimensions of the latter while two others are defined by the thickness of the sheet stacks. Therefore, basic RGO nanosheets and their stacks are responsible for shungite pores of 2-5 nm in size. Following this line, globules may form pores up to 10 nm while extended aggregates of globules obviously form pores of a few tens nm and bigger. This presentation well correlates with the small-angle neutron scattering data that evidence the presence of two sets of pores in shungite carbon in the range of 2-10 nm and above 100 nm [21]. The inner surface of the pores is mainly carpeted with hydrogen atoms due to which water molecules should be located only in the vicinity of remaining oxygen atoms. Taking together, the multitier of structural elements and various porosity make the fractal structure of shungite carbon good self-consistent.

Shungite carbon is formed in aqueous surrounding and although water molecules do not act as active chemical reactants at the oxidation stage they play a very important role at the reduction [31] as well as in composing shungite as a solid.

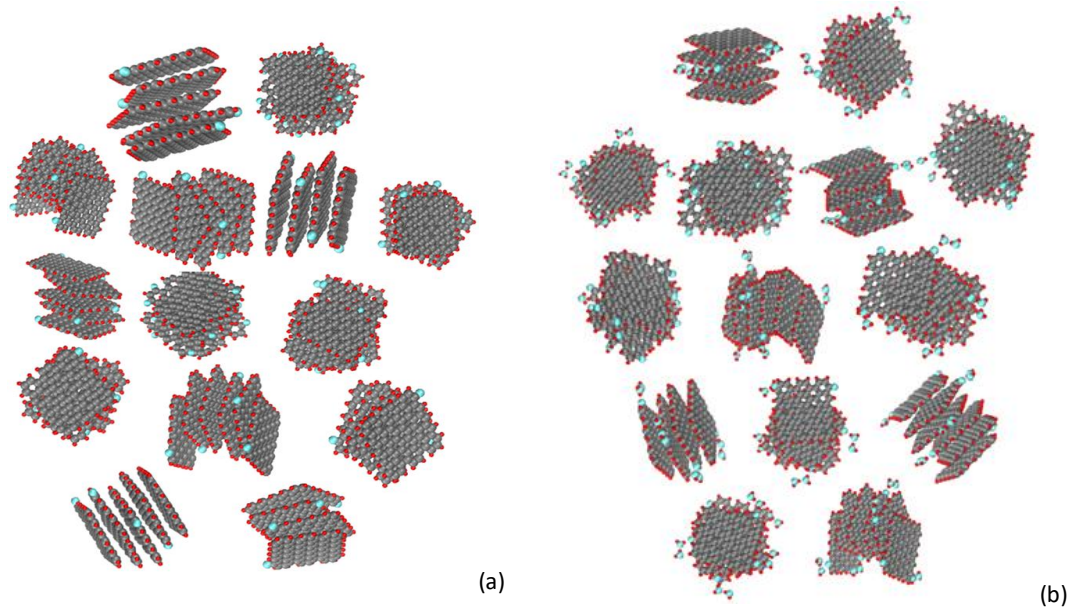

**Figure 5**. Planar view on a model of 'dry' (a) and 'wet' (b) shungite globules consisting of a set of four-, five- and six-layer stacks of 'dry' and 'wet' (5, 5) RGO sheets of the $C_{66}O_1H_{21}$ chemical composition voluntarily located and oriented in space. Linear dimensions along the vertical and horizontal are of ~6 nm.

Firstly, the slow rate of reduction evidently favors the accumulation of RGO nanosheets during a long shungite geological story. Should not exclude also a possible chemical modification of the sheets framing due to their long stay in hot water. Secondly, water molecules fill the pores, helping to strengthen the framework of fractal shungite carbon. These processes, when taken together, have led to the creation of unique natural pantry of nanoscale reduced graphene oxide.

## 5. Conclusive remarks

The main object of the discussions presented in the current paper concerns technical graphene that is under extreme request for low-performance applications of graphene. Technical graphene is a joint name of a large number of varied FGM chemicals a leading role among which belongs to RGOs. Among the latter used in practice, synthetic products take the main place, while a natural one, known as shungite carbon, can do the same work. Two main issues related to technical graphene are considered in the paper: 1) the multi-stage character of the reduction of graphene oxides responsible for a large variety of chemical composition and structure of final RGO products and 2) the oxidation and reduction of graphene lamellae, occurred under natural conditions, leading to the formation of shungite carbon.

Two factors, tightly connected with the coupling energy of OCGs attached to graphene carbon skeleton, govern multistage character of GO reduction: i) the presence of two spatial zones of graphene greatly differing by chemical activity and ii) large-amplitude waving character of the coupling energy depending on the site where an OCG is taken off. The first factor strongly distinguishes circumference and basal plane of a bare graphene molecule in favor of the latter while the second makes the reduction in the circumference area multistage depending on the reaction protocol limiting desorption process by energy. The multistage character of the reduction as a whole explains a large variety of chemical composition of produced RGOs.

The understanding of the discussed peculiarities of the GO reduction allowed throwing light on the formation of RGOs under natural conditions. It was shown that RGOs of shungite carbon were formed in the course of oxidation/reduction/hydrogenation reactions that govern chemical modification of the pristine graphene lamellae. The first two reactions work simultaneously but serving different purposes: oxidation terminates the growth of pristine graphene lamellae thus determining their size, while reduction strengthens the tendency and consequently releases the oxygenated nanosheets from OCGs located through over basal plane and at the sheets perimeter. The emptied sites in the circumference are filled with hydrogen atoms thus inhibiting high reactivity of edge atoms and finally stabilizing shungite RGO as final product. This conclusion is well consistent with shungite carbon empirical data, the main of which are related to exhibiting (1) ~1 nm planar-like RGO sheets as basic structural elements of the macroscopic shungite structure; and (2) low contents of remaining oxygen and hydrogen in the most carbon-pure shungite samples. The existence of this natural product opens up large prospects of low-performance applications based on using technical graphene.

**Acknowledgement**

The author is deeply thankful to N.Rozhkova, E.Golubev and N.Popova for stimulating and fruitful discussions.